\documentclass{elsart}

\usepackage{graphicx}

\usepackage{amssymb}

\begin{document}

\begin{frontmatter}

\title{Stoichiometry control of magnetron sputtered Bi$_2$Sr$_2$Ca$_{1-x}$Y$_x$Cu$_2$O$_y$ (0$\le$x$\le$0.5) thin
film, composition spread libraries: Substrate bias and gas density
factors.}
\author{R. J. Sanderson and}
\author{K. C. Hewitt\corauthref{cor1}}
\corauth[cor1]{Corresponding author.} \ead{Kevin.Hewitt@Dal.ca}
\ead[url]{http://fizz.phys.dal.ca/$\sim$ hewitt}
\address{Dalhousie University, Department of Physics and Atmospheric Science, 6300 Coburg Road, Halifax NS Canada B3H 3J5}

\begin{abstract}
A magnetron sputtering method for the production of thin-film
libraries with a spatially varying composition, x, in
Bi$_2$Sr$_2$Ca$_{1-x}$Y$_x$Cu$_2$O$_y$ (0$\le$x$\le$0.5) has been
developed. Two targets with a composition of
Bi$_2$Sr$_2$YCu$_2$O$_{8.5 + \delta}$ and
Bi$_2$Sr$_2$CaCu$_2$O$_{8 + \delta}$ are co-sputtered with
appropriate masks.  The target masks produce a linear variation in
opposite, but co-linear radial direction, and the rotation speed
of the substrate table is sufficient to intimately mix the atoms.
EDS/WDS composition studies of the films show a depletion of Sr
and Bi that is due to oxygen anion resputtering. The depletion is
most pronounced at the centre of the film (i.e. on-axis with the
target) and falls off symmetrically to either side of the 75 mm
substrate.  At either edge of the film the stoichiometry matches
the desired ratios. Using a 12 mTorr process gas of argon and
oxygen in a 2:1 ratio, the strontium depletion is corrected.  The
bismuth depletion is eliminated by employing a rotating carbon
brush apparatus which supplies a -20 V DC bias to the sample
substrate.  The negative substrate bias has been used successfully
with an increased chamber pressure to eliminate the resputtering
effect across the film.  The result is a thin film composition
spread library with the desired stoichiometry.
\end{abstract}

\end{frontmatter}

\section{Introduction}
\label{intro}

It is widely accepted that oxygen has to be added to the
sputtering gas for the deposition of Bi$_2$Sr$_2$CaCu$_2$O$_{8 +
\delta}$ (Bi-Ca-2212) thin films
\cite{park95,hammond89,westerheim91,morris89}. The drawback to
using oxygen in the sputtering gas is a pronounced resputtering of
the deposited film, as found in Bi-Ca-2212 \cite{grace92,grace91}
and YBa$_2$Cu$_3$O$_{7-\delta}$
\cite{xu94,nathan99,blue91,ballentine91,klien91} superconducting
systems. Oxygen anions derived from the sputtering gas and target
material are the source of this resputtering effect
\cite{cuomo78}. The anion is accelerated away from the negatively
charged target, and if the energy is large enough, it will escape
the magnetron's magnetic field. The energetic anions sputter the
deposited film, in a manner than preferentially depletes atoms
with large scattering cross section. This resputtering effect
produces the undesirable effect of changing the composition of the
film relative to the target.

There have been several reported methods to reduce the
resputtering phenomenon in these materials.  The most common
approach is to use an off-axis deposition method, where the
substrate is not directly across from the target.  Rather, the
substrate is placed at an angle $<$ 90$^o$ relative to the axis of
the cylindrical target.  It has been found that the energy of
oxygen anions that strike the film is reduced
\cite{park95,grace92,grace91,xu94,nathan99,blue91,ballentine91,klien91}.
Off-axis sputtering greatly reduces the deposition rate, which is
not desirable.  To further reduce resputtering, it has been shown
that increasing the sputtering gas pressure can increase the
anion-gas collisions, reducing their energy or deflecting them
away from the film.  Increasing the gas pressure usually reduces
the target bias. The energy of the anions is thereby decreased;
however, the increased gas pressure also decreases the film
deposition rate due to target atom-gas collisions.  A third method
to compensate for resputtering is to alter the target
stoichiometry to account for the non-uniform depletion of elements
\cite{grace92,xu94,blue91}; this is sometimes not possible as the
required target composition may not be an equilibrium phase. A
final method uses a negative bias shield or substrate bias to
deflect the anions away from the deposited film
\cite{park95,westerheim91}.

Physical vapor deposition has been used since 1965 to explore
phase diagrams using a composition-spread approach
\cite{kennedy65,hanak70,xiang95,vanDover98,chang99,xiang99}.  In
this technique, a large range of compositions is deposited on a
single substrate in a single deposition run.  Our group has
established a composition spread approach which uses magnetron
sputtering to prepare a continuous variation of composition across
a 75 mm substrate \cite{dahn02}.

Therefore, to develop a magnetron sputtering method for the
production of cuprate superconductor thin-film samples with a
spatially varying composition, x, in
Bi$_2$Sr$_2$Ca$_{1-x}$Y$_x$Cu$_2$O$_y$ (0$\le$x$\le$0.5),
resputtering effects have to be addressed as it has a detrimental
effect on the composition of the deposited film. An apparatus had
to be designed to bias the substrates while affixed to a rotating
substrate table.  The novel aspect of this paper is the
introduction of a rotation-compatible substrate bias apparatus
that uses an existing magnetron to supply a constant -20 V bias to
the substrates in order to prepare a cuprate superconductor
composition spread library.

\section{Experiment and Results}
\label{exp}

The composition spread approach uses simultaneous, magnetron
sputtering of two targets, e.g. A and B.  Masks are placed over
each target to produce a linear variation in the mass deposited
onto a water-cooled, rotating substrate as described in more
detail in reference \cite{dahn02}.  When the linear variation is
in opposite, but co-linear radial direction, and the rotation
speed (15 revolutions per minute) is sufficient to intimately mix
the atoms, a film A$_{1-x}$B$_x$ (0$\le$x$\le$1) is produced
radially across a 7.5 cm doughnut on a 43 cm diameter substrate
table - the doughnut is centered at a radius of 13.5 cm from the
substrate centre. To produce a film
Bi$_2$Sr$_2$Ca$_{1-x}$Y$_x$Cu$_2$O$_y$ (0$\le$x$\le$1), one must
therefore co-sputter two targets:  A = Bi$_2$Sr$_2$CaCu$_2$O$_{8 +
\delta}$ (Bi-Ca-2212) and B = Bi$_2$Sr$_2$YCu$_2$O$_{8.5 +
\delta}$(Bi-Y-2212).

The powders used to synthesize the targets are made through the
same three-stage solid-state reaction sequence.  To synthesize
Bi-Ca-2212, powders of Bi$_2$O$_3$ (99.975 \%, Alfa-Aesar),
SrCO$_3$ (99.99 \%, Alfa-Aesar), CaO (99.95 \%,Alfa-Aesar) and CuO
(99.7 \%,Alfa-Aesar) are measured in the appropriate
stoichiometric ratios and ground together for two hours using an
agate auto grinder. The mixture is first reacted in air at 800
$^o$C  for 12 hours (slow heat/ slow cool at 4$^o$C/min), to
calcinate the powders. After the first reaction, the powder is
ground for 2 hours. The next two reactions in air are at 870
$^o$C. Between these two 870$^o$C reactions the powder is ground
for 2 hours. To produce Bi-Y-2212,  Y$_2$O$_3$ (99.9 \%,
Alfa-Aesar) is used instead of CaO. Once the target powders are
synthesized, they are pressed into pucks and hardened before use.
To accomplish this, the powder is ground manually with an agate
mortar and pestle and sifted through a 70 $\mu$m sieve. Next,
approximately 40 g of the sieved powder is pressed into a 5.08 cm
(2") diameter by 0.5 cm thick disc using a pressure of 8500 psi.
The disc is sintered to harden the target and reduce porosity. The
sintering temperature was chosen to be close to the melting
temperature for each of the two targets (Bi-Ca-2212 - 875 $^o$C
and Bi-Y-2212 - 935 $^o$C).

\begin{figure}[h!]
\includegraphics{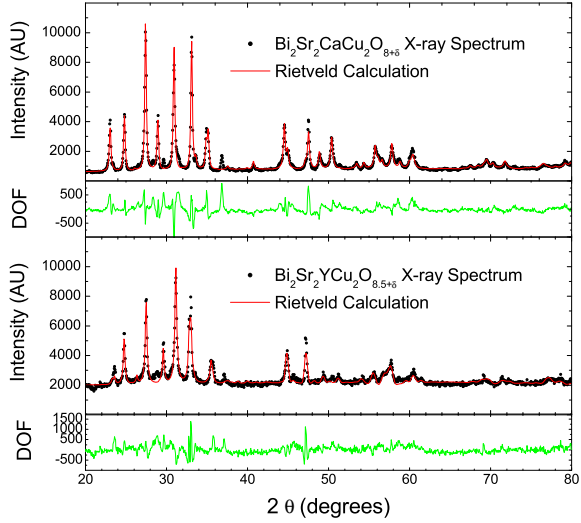}
\vspace{0.1in} \caption{Rietica analysis fit of the XRD patterns
of Bi-Y-2212 and Bi-Ca-2212 target materials. DOF stands for
difference of fit.} \label{XRD}
\end{figure}

Figure~\ref{XRD} shows x-ray spectra of the Bi-Ca-2212 and
Bi-Y-2212 target materials collected with a Siemens D500 x-ray
spectrometer using Cu K$_\alpha$ x-rays.
Using the Bbmb (D$_{2h}^{20}$) space group (non-conventional
setting of Cccm), the powder x-ray spectrum for each target was
fit with Rietica V1.62 - a Rietveld analysis program.
Figure~\ref{XRD} shows the x-ray diffraction spectra, as well as
the fit for each material, and Table
~\ref{rietica:parameters-Bi-Ca-2212} and
~\ref{rietica:parameters-Bi-Y-2212} shows the refined parameters.
\begin{table}[h!!tb]
\caption{Rietica fit parameters for Bi-Ca-2212 (Bi/Sr means Bi in
Sr sites and Ca/Sr means Ca in Sr sites).}
\begin{tabular}{ccccc}
\hline
Space group : Bbmb(D$_{2h}^{20}$) &&&& \\
a = 5.405(3) {\AA}, & b = 5.402(2) {\AA}, & c = 30.804(2) {\AA}& \\
Goodness of fit:  5.32 &&&& \\
 \hline
Atom & x & y & z & Occ. \\ \hline \hline
  Bi & 0.5 & 0.515(2) & 0.2425(1) & 1.05(7)   \\ \hline
  Sr & 0.0 & 0.1394(6) & 0.2448(3) & 0.91(1)   \\ \hline
  Ca & 0.0 & 0.25 & 0.25 & 0.91(1)   \\ \hline
  Cu & 0.5  & 0.1947(4) & 0.2370(5) & 1.04(7)  \\ \hline
  Bi/Sr & 0.0 & 0.1394(6) & 0.2448(3) & 0.053(4) \\ \hline
  Ca/Sr & 0.0 & 0.1394(6) & 0.2448(3) & 0.036(1) \\ \hline
  O(1) & 0.0  & 0.1695(7) & 0.057(9) & 1.7(5) \\ \hline
  O(2) & 0.25  & 0.1193(1) & 0.2974(3) & 1.1(1) \\ \hline
  O(3) & 0.0  & 0.0287(5) & 0.1374(4) & 0.5(8) \\ \hline
  \hline
\end{tabular}
\label{rietica:parameters-Bi-Ca-2212}
\end{table}

\begin{table}[h!!tb]
\caption{Rietica fit parameters for Bi-Y-2212.}
\begin{tabular}{ccccc}
\hline
Space group : Bbmb(D$_{2h}^{20}$) &&&& \\
a = 5.45(9) {\AA}, & b = 5.428(5) {\AA}, & c = 30.185(2) {\AA} &&\\
Goodness of fit:  7.28 &&&& \\
 \hline
Atom & x & y & z & Occ. \\ \hline \hline
  Bi & 0.5 & 0.503(9) & 0.2763(1) & 0.88(0)   \\ \hline
  Sr & 0.0 & 0.1354(4) & 0.2444(3) & 1.34(1)   \\ \hline
  Y & 0.0 & 0.25 & 0.25 & 0.6882(6)   \\ \hline
  Cu & 0.5  & 0.195(8) & 0.232(2) & 0.7293(4)  \\ \hline
  O(1) & 0.0  & 0.152695(7) & 0.0 & 1.328(8) \\ \hline
  O(2) & 0.25  & 0.1892(2) & 0.5 & 1.8(9) \\ \hline
  O(3) & 0.5  & 0.1605(6) & 0.2801(4) & 1.2(6) \\ \hline
  \hline
\end{tabular}
\label{rietica:parameters-Bi-Y-2212}
\end{table}
The fits to the spectra demonstrate that the space group Bbmb is a
reasonable choice for the Bi-2212 phase. Goodness of fit (GOF)
values for the Bi-Ca-2212 fit of 5.32 and 7.28 for the Bi-Y-2212
system are large. This large GOF is due to the fact that the
orthorhombic space group is well known to represent an average
structure which does not account for the incommensurate modulation
along the b-axis of 4.7b.

The change in unit cell size is consistent with previous reports
\cite{hsu00}: the a-axis lattice parameter expands from 5.41 {\AA}
to 5.46 {\AA}, b expands from 5.40 {\AA} to 5.43 {\AA} and c
contracts from 30.80 {\AA} to 30.18 {\AA}.  Overall, the volume
contracts by 0.6 \% upon substitution of Y for Ca in accordance
with previous reports which find a volume reduction of 0.7 \%
\cite{hsu00}.

\begin{figure}[h!]
\includegraphics{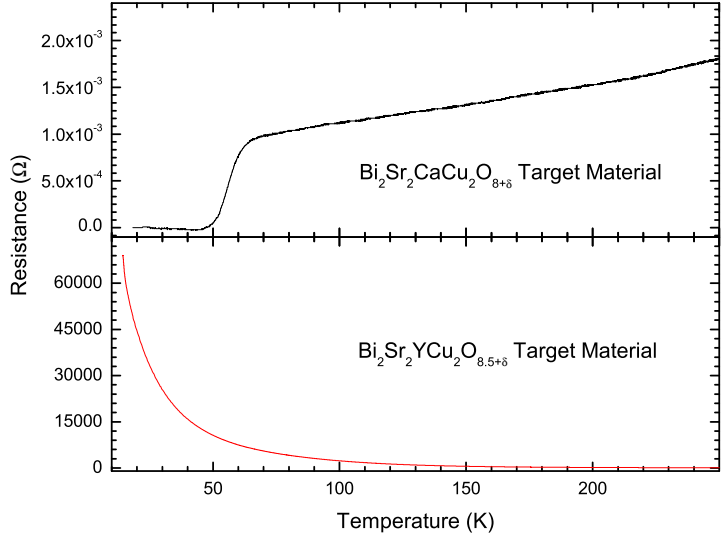}
\vspace{0.1in} \caption{Low temperature four-probe resistance
measurements of Bi-Y-2212 and Bi-Ca-2212 target materials.}
\label{transportpropertiesoftargets}
\end{figure}

To confirm the electrical properties of each material, low
temperature resistance measurements were made.  Each target
material was pressed into a 1.2 cm diameter x 0.5 cm pellet and
sintered.  Silver contact pads were evaporated onto the pellet to
improve conduction to the sample.  Four Cu wires were affixed with
silver epoxy to the pellet on top of the silver pads, to perform a
4-probe resistance measurement.  The pellets were mounted into a
Janis CCS-450 cold helium gas compressor cryostat system, and
cooled as a constant current of 100 mA was applied and the
subsequent voltage measured.  Using V/I, the resistance versus
temperature plots for the two materials are shown in Figure
~\ref{transportpropertiesoftargets}.
It can be clearly seen that the Bi-Y-2212 material behaves as an
insulator, while the Bi-Ca-2212 material shows a T$_{c,onset}
\approx$67 K.

The film deposition apparatus is a Corona Vacuum Coaters V-37
sputtering system equipped with 5 magnetrons configured in a
side-sputtering arrangement, where the substrate is directly
across (5.5 cm) from the target.  To power the magnetrons either
an Advanced Energy MDX-1K DC supply or a combination of an
Advanced Energy RFX-600 generator and RTX-600 tuner is chosen for
the Bi-Ca-2212 (conductor) and Bi-Y-2212 (insulator) targets,
respectively. Typical sputtering conditions are shown in
Table~\ref{sputtering conditions}.
\begin{table}[h!!tb]
\caption{Typical sputtering conditions.}
\begin{tabular}{ccc}

 \hline
Target & Bi-Ca-2212 & Bi-Y-2212  \\ \hline \hline
 Sputtering power& 70 W DC & 109 W RF    \\ \hline
  Base pressure (Torr) & \multicolumn{2}{c}{4.0 x 10$^{-7}$}    \\ \hline
  Deposition time & \multicolumn{2}{c}{60 hours}  \\ \hline
  Deposition thickness & \multicolumn{2}{c}{1.2 $\mu$m}  \\ \hline
  \hline
\end{tabular}
\label{sputtering conditions}
\end{table}
Energy dispersive spectroscopy (EDS) measurements were made using
a JEOL JXA-8200 Superprobe equipped with a Noran energy detector
(0.133 keV energy resolution). A 15kV electron beam with a 20 nA
current is used to analyze a 10 $\mu$m spot. The film is deposited
onto polished Al or Si-wafer substrates for composition analysis.
Polished Al was chosen due to partial x-ray peak overlap between
the K$_\alpha$ and K$_\beta$ x-rays of Si (1.74 and 1.83 keV
respectively) and the L$_\alpha$ and L$_\beta$ x-rays of Sr (1.81
and 1.87 keV respectively). Polished Si wafers provide a better
surface for determination of Bi, Ca and Cu.   In addition, the
film is deposited onto single crystal (100) MgO substrates,
through a slotted aluminum mask (0.5 mm slots separated by 1 mm
across the 75 mm substrate), for post annealing studies.

For all deposited films, X-ray diffraction (XRD) spectra were
collected using an Inel CPS-120 with a curved position sensitive
detector. The Cu-K$_{\alpha_1,\alpha_2}$ x-ray beam is incident
upon the sample at approximately 6$^o$ and the curved position
sensitive detector collects all scattered x-rays from 2$\theta$ =
3 $^o$ - 119$^o$. Collection time for the XRD spectra was
typically 1800 seconds. The Inel and the JEOL both have computer
controlled sample stages to scan samples precisely and
efficiently.

Using energy dispersion spectroscopy (EDS) the normalized
composition of Bi and Sr across the deposited film from three runs
is shown in Figure~\ref{EDS-pressure-gas ratio effect}.
\begin{figure}[h!]
\includegraphics{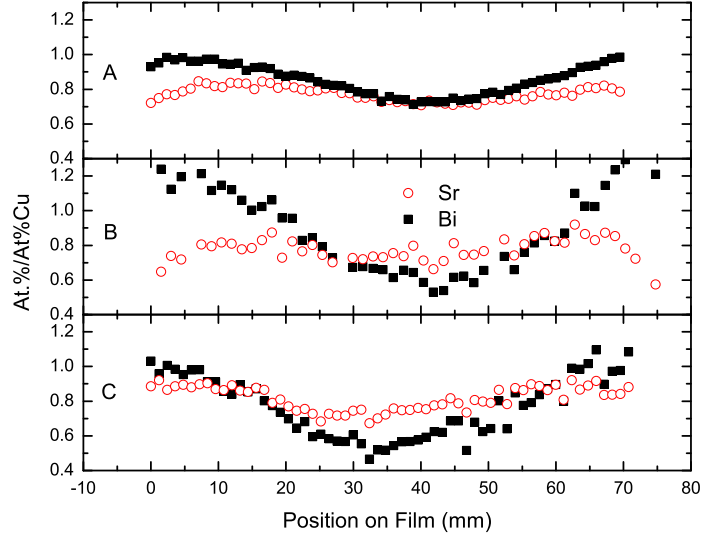}
\vspace{0.1in} \caption{EDS compositional analysis of Bi (filled
square) and Sr (empty circle) as per position along the film. Data
has been normalized to the atomic percent Cu present in the film.
A) SPI102, B) SPJ098 and C) SPI131.} \label{EDS-pressure-gas ratio
effect}
\end{figure}
Panels A, B and C correspond to sputtering conditions as listed in
Table~\ref{sputtering conditions - gas pressure effect}.
\begin{table}[h!!tb]
\caption{Sputtering gas conditions for results presented in
Fig.~\ref{EDS-pressure-gas ratio effect}.}
\begin{tabular}{cccc}

 \hline
Fig. \ref{EDS-pressure-gas ratio effect} panel & A (SPI102) & B
(SPJ098) & C (SPI131) \\ \hline \hline
 Base pressure (Torr)& 4.57 x 10$^{-7}$ & 4.92 x 10$^{-7}$ & 3.52 x 10$^{-7}$    \\ \hline
Chamber pressure (Torr)& 5.71 x 10$^{-3}$ & 12.0 x 10$^{-3}$ &
13.2 x 10$^{-3}$    \\ \hline
 O$_2$:Ar flow rate (sccm) & 2:4 & 5:10 & 1:13   \\ \hline
  O$_2$:Ar ratio & 1:2 & 1:2 & 1:13   \\ \hline
  \hline
\end{tabular}
\label{sputtering conditions - gas pressure effect}
\end{table}

The data shown in Figure~\ref{EDS-pressure-gas ratio effect}
illustrates the effect that both oxygen concentration and
sputtering pressure have on resputtering.  Panel A shows
resputtering in both Bi (30 \% depletion at the center of film
relative to edge) and Sr (13 \% depletion at the center of film
relative to edge) with a 1:2 = O$_2$:Ar ratio. Panel B shows there
is reduced resputtering of Sr ($<$1 \%) as the O$_2$:Ar ratio is
kept constant, but the pressure increased.  However, as seen in
the data of panel B, Bi is still resputtered (42 \% depletion at
the center of film relative to edge).  Panel C shows the data
collected from a sputtering run with the working pressure
comparable to panel B, but the O$_2$:Ar ratio decreased to 1:13.
The degree of resputtering is comparable to that which is present
in panel B (40 \% depletion at the center of film relative to
edge), but there was significant damage to the Bi-Ca-2212 target.
This damage may be due to oxygen depletion on the surface of the
target, and has not been present in runs with increased O$_2$:Ar
ratios. Of the data presented in Figure \ref{EDS-pressure-gas
ratio effect}, panel B shows the most desirable characteristics.

\begin{figure}[h!]
\includegraphics{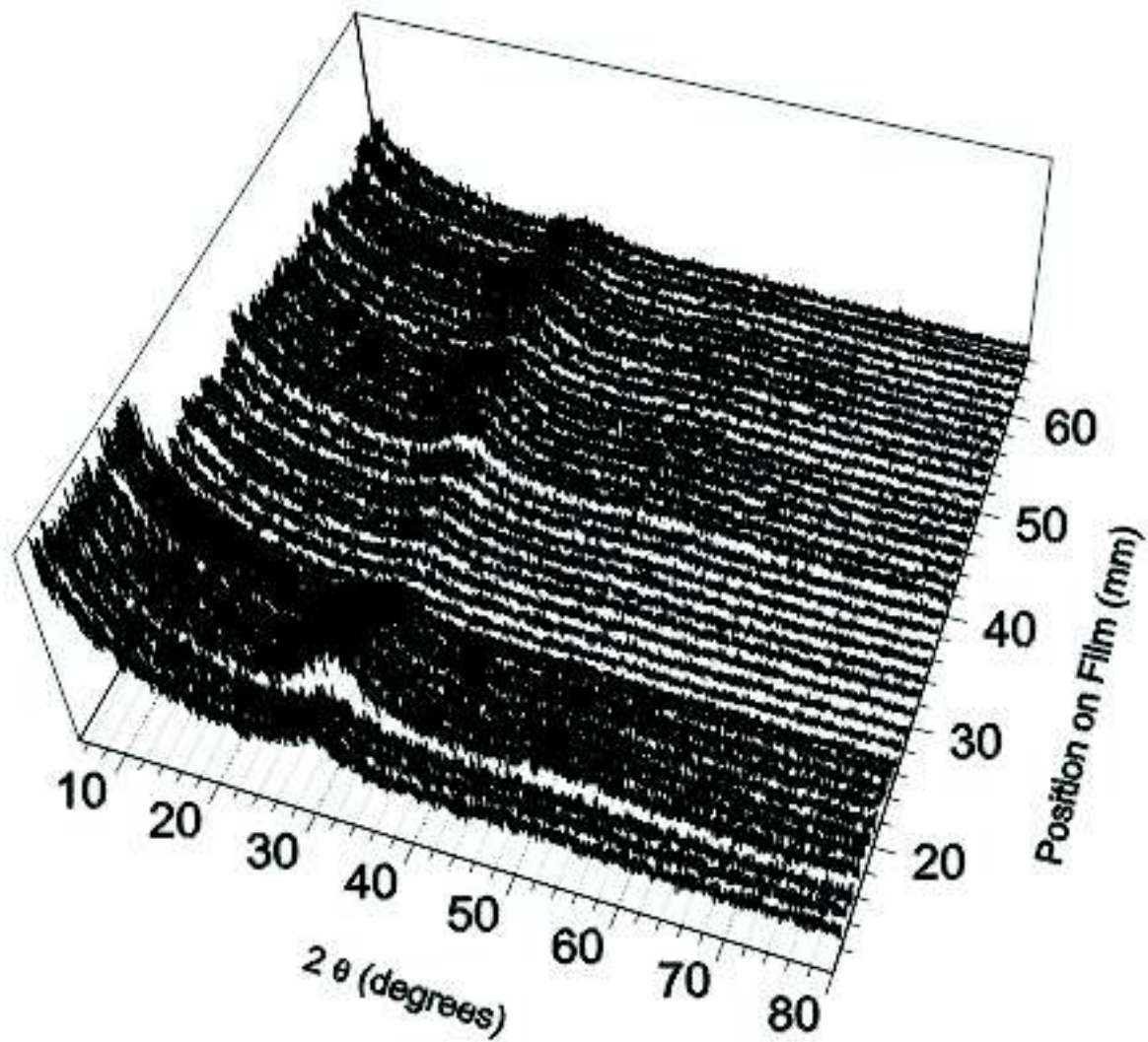}
\vspace{0.1in} \caption{XRD spectra of the as-deposited film
(SPJ098) as a function of position along the substrate.}
\label{XRD-as deposited}
\end{figure}

Figure~\ref{XRD-as deposited} shows XRD spectra of the
as-deposited films from the deposition that produced the
composition shown in Figure~\ref{EDS-pressure-gas ratio effect},
panel B as a function of position along the film.
The presence of broad peaks shows the amorphous nature of the
films.  As such, the films must be annealed to crystallize the
Bi-2212 phase.

The as-deposited film whose EDS data are shown in
Figure~\ref{EDS-pressure-gas ratio effect} (panel B) was post
annealed in a covered alumina crucible. According to Rubin
\cite{rubin94}, the deposited films can develop Bi deficiency
during annealing.  To compensate, they added Bi$_2$O$_3$ powder to
enrich the atmosphere with Bi, since Bi$_2$O$_3$ has a lower vapor
pressure than the BiO in the films \cite{rubin94}. Adapting the
annealing regime from Tsukamoto \cite{tsukamoto93} and using
Rubin's suggestion, the films were annealed as follows. The
furnace was heated to 780 $^o$C at which point the film,
previously placed in a covered alumina crucible with powdered
Bi$_2$O$_3$, was inserted. The crucible was introduced to the
furnace for 0.5 hours, after which it was removed from the furnace
and air quenched to room temperature. The furnace was then heated
to 860 $^o$C and the samples were again inserted into the hot
furnace for 6 hours, and then removed.

\begin{figure}[h!]
\includegraphics{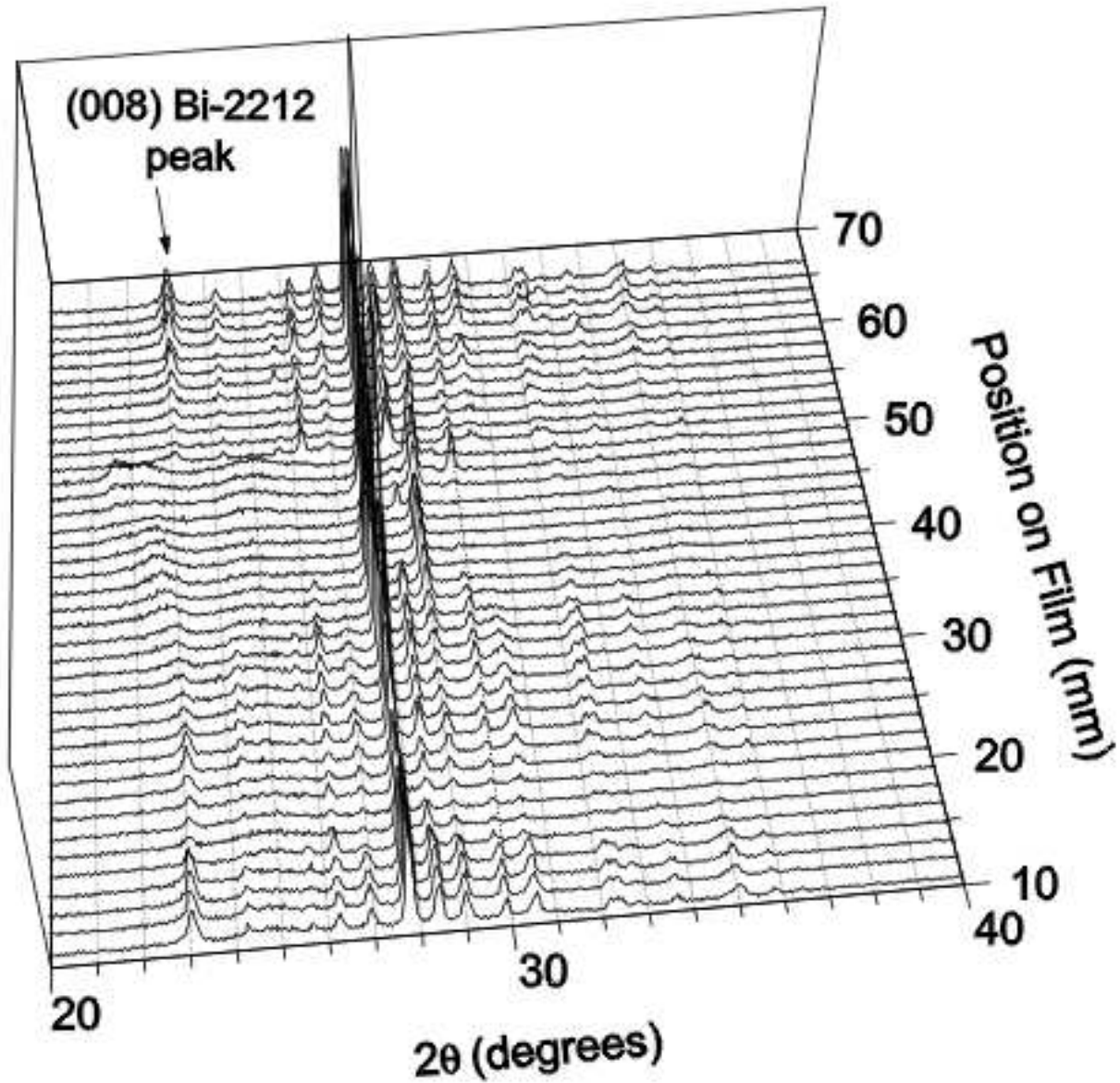}
\vspace{0.1in} \caption{XRD of the post-annealed film (SPJ098) as
a function of position along the substrate. The (008) peak of the
Bi2212 structure is indicated by an arrow.} \label{XRD-annealed}
\end{figure}

\begin{figure}[h!]
\includegraphics{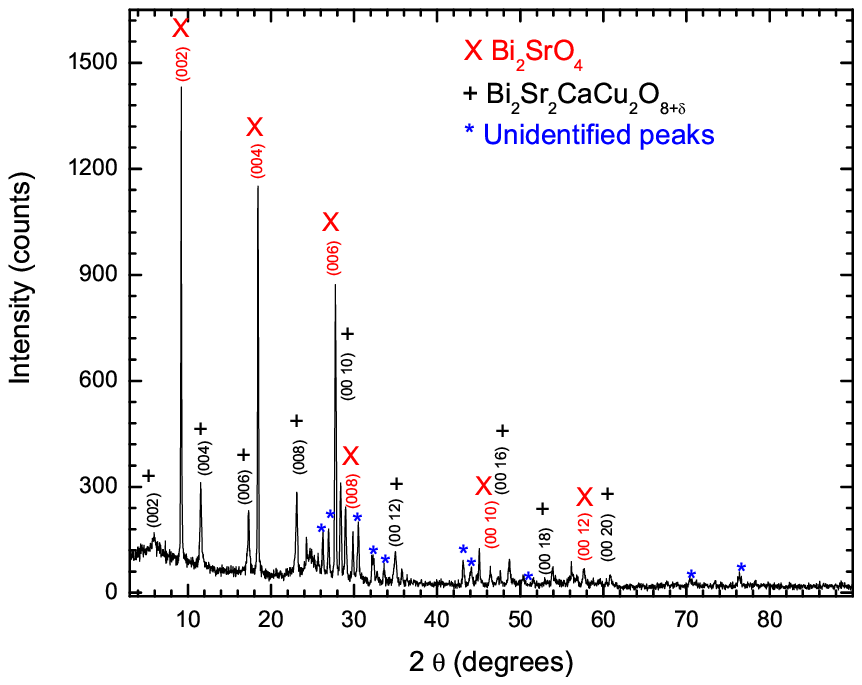}
\vspace{0.1in} \caption{XRD spectra of the post-annealed film
(SPJ098) at a position of 10.4 mm (see Figure \ref{XRD-annealed}).
Shown are the ($\it{hkl}$) indicies of the peaks of the two
predominant phases, Bi$_2$SrO$_4$ and
Bi$_2$Sr$_2$CaCu$_2$O$_{8+\delta}$.} \label{XRD-10mm}
\end{figure}

X-ray diffraction patterns of the annealed film deposited on MgO
are shown in Figure~\ref{XRD-annealed}.
The XRD spectra reveal the presence of two major phases - the
desired Bi2212 phase and a second, more dominant, Bi$_2$SrO$_4$
phase. Figure~\ref{XRD-10mm} shows the XRD spectra corresponding
to a point 10.4 mm from the inner edge of the film, and the hkl
indicies of the two phases.
It is possible that the Bi$_2$SrO$_4$ phase is induced from the
presence of Bi$_2$O$_3$ in the anneal atmosphere \cite{rubin94},
though there is no firm evidence of this. Figure ~\ref{XRD-10mm}
also reveals that the film exhibits epitaxial growth, since only
($\it{hkl}$)=(00$\it{l}$) X-ray peaks are observed. Thus, the
films are oriented with the c-axis perpendicular to the plane of
the substrate - a very desirable feature. The intensity of the
(008) peak of the Bi2212 phase diminishes as the film progresses
toward the center, and begins to reappear and increase in
intensity past the center.

\begin{figure}[h!]
\includegraphics{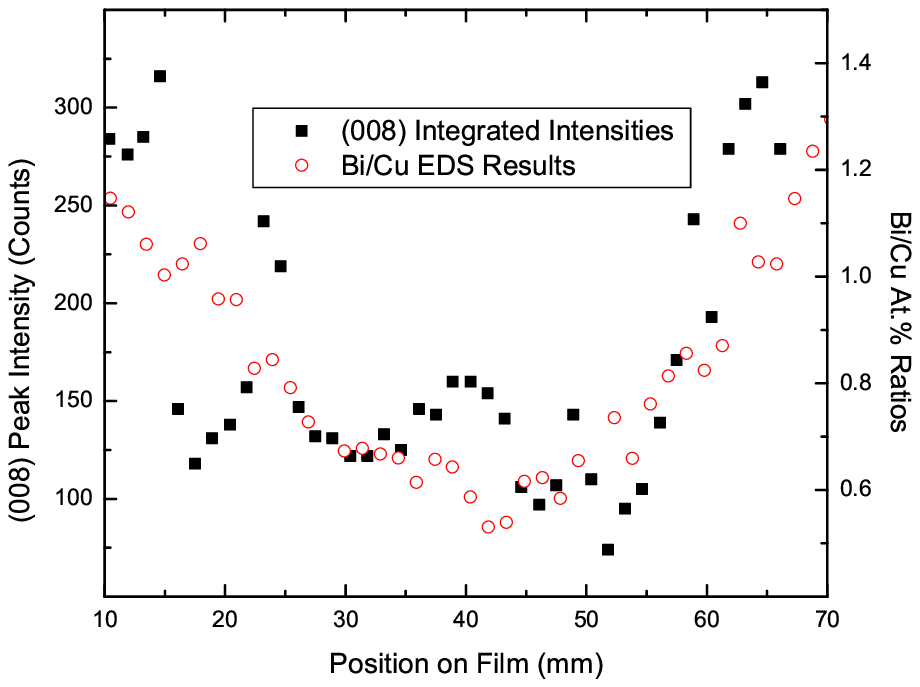}
\vspace{0.1in} \caption{Integrated intensity of the (008) Bi2212
peak and EDS results from figure \ref{EDS-pressure-gas ratio
effect}, panel B plotted as per position along the deposited film
(SPJ098).} \label{008 integration and Bi-Cu EDS comparison}
\end{figure}

Figure \ref{008 integration and Bi-Cu EDS comparison} shows the
correlation of (008) peak area with the Bi/Cu ratio as seen in
Figure \ref{EDS-pressure-gas ratio effect}, panel B.
It is obvious that deposition of the correct Bi concentration is
needed to create the desired Bi-2212 phase.

\begin{figure}[h!]
\includegraphics{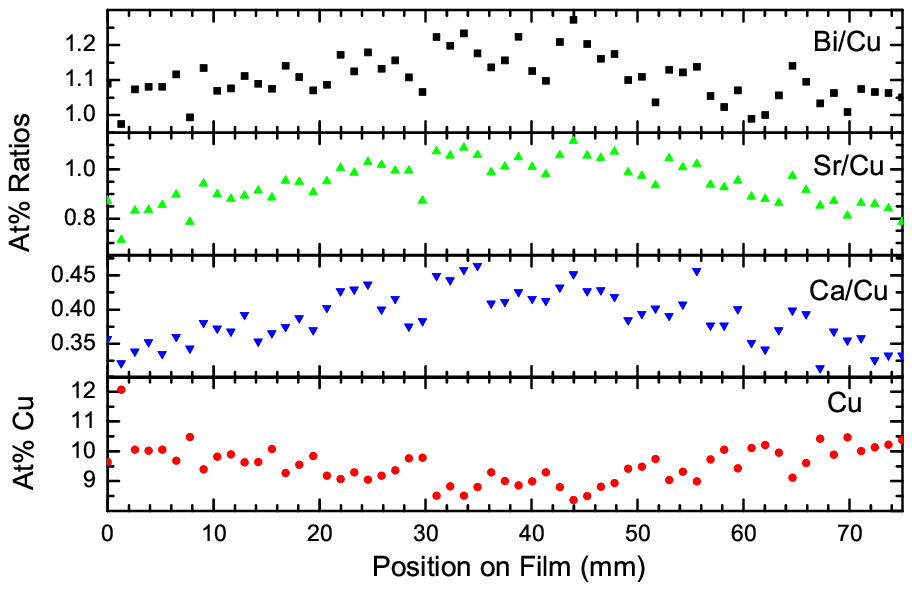}
\vspace{0.1in} \caption{EDS data of a Bi-Ca-2212 deposition
(SPJ113). The peaks in the normalized atomic ratios is from the
uncharacteristic depletion of Cu in the center of the film.}
\label{EDS - pure Bi2212 deposition}
\end{figure}

In an attempt to produce a pure Bi-2212 phase, a deposition was
made with just the Bi-Ca-2212 target.  This run was carried out
with the same conditions as listed in Table ~\ref{sputtering
conditions - gas pressure effect}.  Single crystal (100) MgO
substrates were placed on the edge of the film (over the last 10
mm of the sputtering track), without any physical masks. A film of
approximately 1.5 $\mu$m was deposited.  EDS of the film is shown
in Figure~\ref{EDS - pure Bi2212 deposition} where it is seen that
the normalized data for Bi, Sr and Ca appear to have a peak in the
center.
Further inspection revealed an unexpected deficiency of Cu at the
center which increases the normalized atomic ratio of Bi in the
center.  However, near the edges the stoichiometry matches the
desired ratios. This problem with Cu deficiency has not resurfaced
in numerous depositions.

The film was subject to an annealing regime consisting of a quick
ramp (25 $^o$C/min) to 780 $^o$C for one hour followed by a second
identical quick ramp to 860 $^o$C for three hours. As shown in
Figure ~\ref{XRD of epitaxial Bi2212 film} the result is a
partially crystalline phase.
\begin{figure}[h!]
\includegraphics{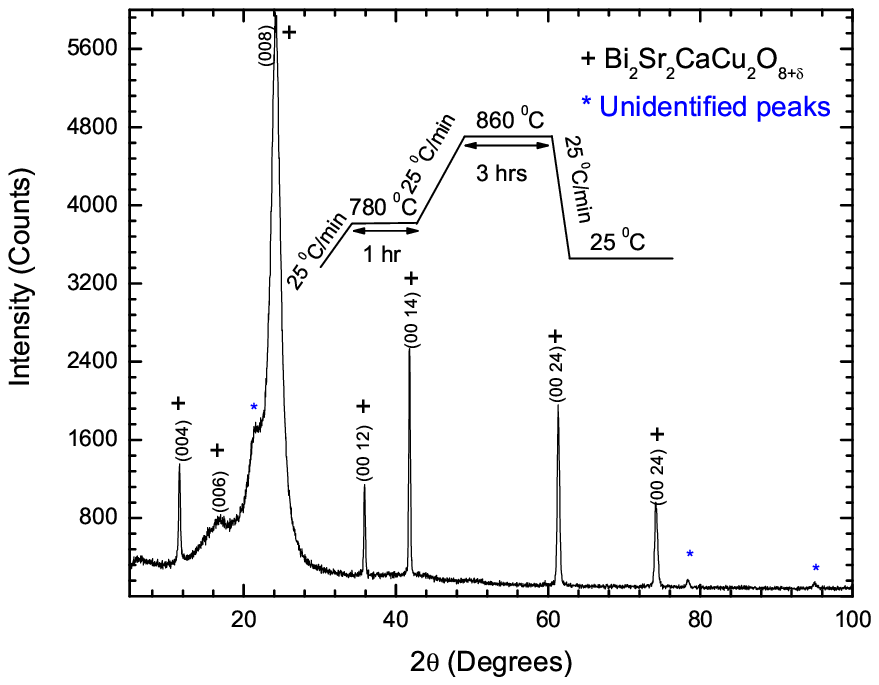}
\vspace{0.1in} \caption{XRD of a post annealed film (SPJ113)
segment (0-10mm) from the deposition whose EDS results are shown
in Figure~\ref{EDS - pure Bi2212 deposition}.  The film was post
annealed as shown in the diagram.} \label{XRD of epitaxial Bi2212
film}
\end{figure}
The sharp peaks in the pattern were identified as (00$\it{l}$)
peaks of the Bi2212 phase (with two weak unidentified peaks).  The
(006) peak is broad demonstrating the partially amorphous nature
of the annealed films. The film would have to undergo a longer
anneal period to fully crystallize. However the lack of the
Bi$_2$SrO$_4$ phase is reassuring.

Though the film composition at the edges are in the desired ratio
and produce the desired phase, there is still a non-stoichiometric
proportion of Bi near the centre of the film. To reduce this
effect, a negative substrate bias had to be constructed to deflect
oxygen anions away from the film. The film is deposited onto a
rotating substrate. Thus one needs to design a rotating contact
brush system.  The substrates are either conductive, or have a
conductive mask, consequently creating a substrate/mask bias is
easily done, if an appropriate electrical contact system can be
designed.  Figure ~\ref{carbon brush apparatus} shows a schematic
of a carbon brush assembly that was manufactured on-site.
\begin{figure}[h!]
\includegraphics{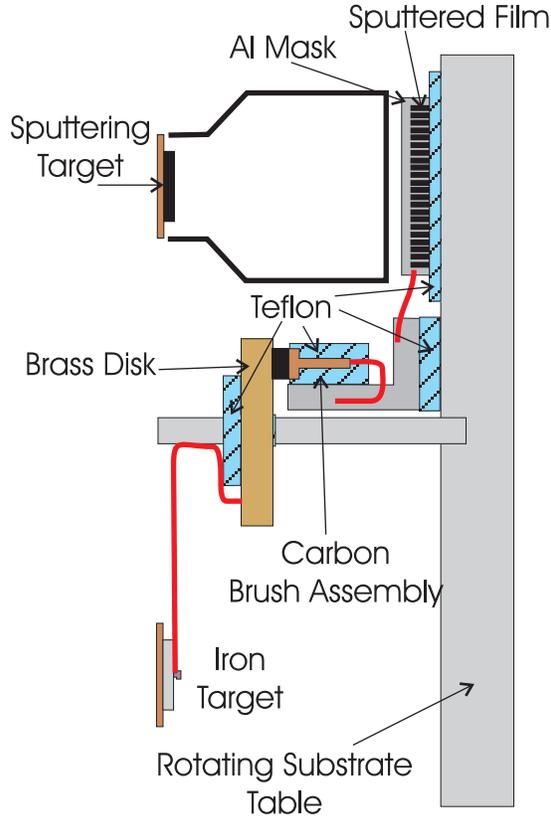}
\vspace{0.1in} \caption{ Schematic of the rotating substrate bias
apparatus.  As the table rotates the carbon brush remains in
contact with the brass disk that has been charged by an available
magnetron.  A bias of -20V is applied to an Al mask to deflect the
oxygen anions, and eliminate resputtering.} \label{carbon brush
apparatus}
\end{figure}
It uses an existing magnetron to supply a negative bias to the
substrate.

As the substrate table rotates, the carbon brush remains in
contact with the brass disk.  The brass disk is fixed to a central
rod which is attached to the back of the sputtering chamber.  An
insulated wire electrically connects the brass disk to a 7 mm
thick iron target. The iron target is thick enough so that the
magnetic field from the magnetron is contained in the iron and
only a small bias is applied, hindering the possibility of the
formation of plasma and the subsequent sputtering of the iron.
During deposition, a negative bias is applied to the iron target,
and through the carbon brush assembly this bias is passed to the
rotating substrate table. Using insulated wires the carbon brush
assembly is connected to conductive Al masks (for MgO) or Al or Si
substrates.  Any negative bias can be applied by the magnetron,
and in this case -20 V DC was chosen.  The -20 V DC bias should be
enough to reduce resputtering when teamed with an increased
working pressure, but not high enough to induce any sputtering of
the sample from any positive ions in the chamber. The brass disk,
carbon brush assembly and the substrate are electrically insulated
from the chamber by Teflon insulators.

Figure ~\ref{EDS demonstration effectiveness of bias} shows EDS
data for a Bi$_2$Sr$_2$Ca$_{1-x}$Y$_x$Cu$_2$O$_y$ (0 $< x < $0.4)
deposition using the rotating substrate bias apparatus with a bias
of -20 V DC, and sputtered with similar conditions to column B in
Table \ref{sputtering conditions - gas pressure effect}.
\begin{figure}[h!]
\includegraphics{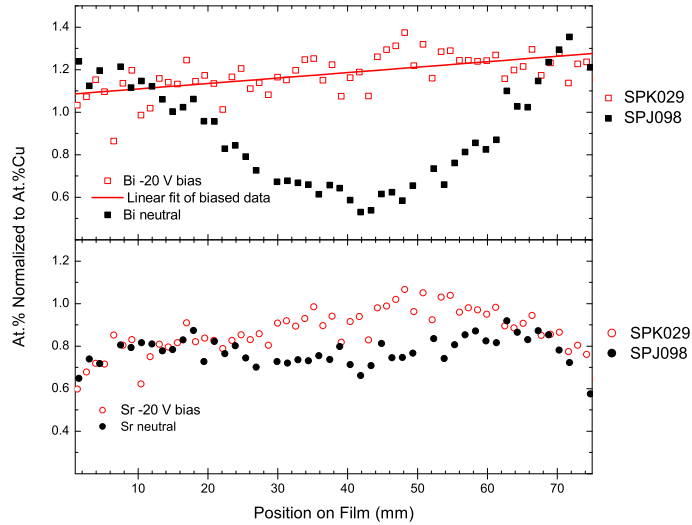}
\vspace{0.1in} \caption{ EDS data from two sputtering runs carried
out under identical sputtering conditions except that in one of
the runs the aluminum substrate was charged with a -20 V DC bias
(open symbols - SPK029) and in the other run the substrates were
left neutral (filled symbols - SPJ098). Films deposited on
polished Al substrates.} \label{EDS demonstration effectiveness of
bias}
\end{figure}
The uniformity of the normalized Bi content across the biased
sample demonstrates the effectiveness of the substrate bias to
eliminate the resputtering effect.  The similarities between the
Sr content in both samples is also reassuring, in that the
composition of the Sr was not greatly affected by the sample bias.

The complete as-deposited EDS profile for co-deposition of
Bi-Y-2212 and the Bi-Ca-2212 targets with a -20 V substrate bias
is shown in Figure~\ref{EDS of composition spread with substrate
bias}.
\begin{figure}[h!]
\includegraphics{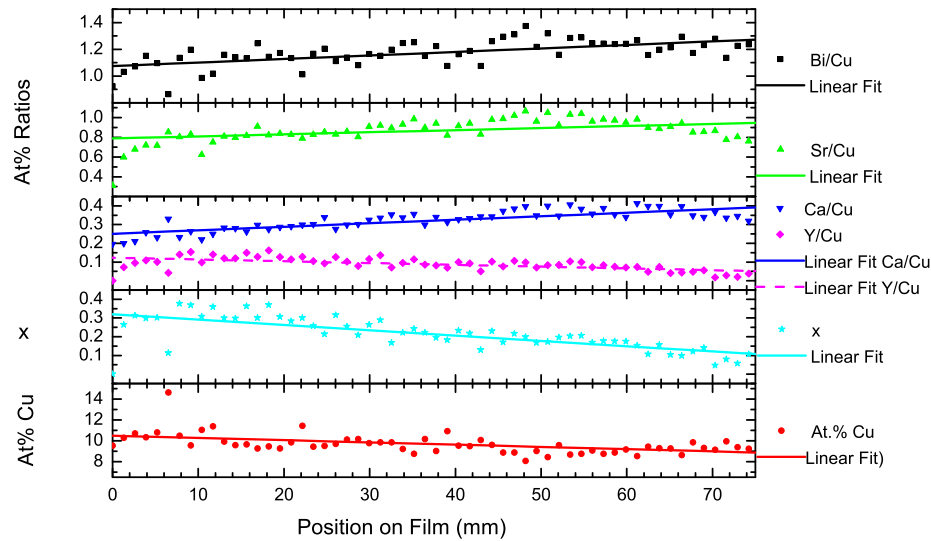}
\vspace{0.1in} \caption{EDS data from a run with both Bi-Y2212 and
Bi-Ca-2212 targets (SPK029). A -20V substrate bias is used.}
\label{EDS of composition spread with substrate bias}
\end{figure}
The levels of Bi, Sr and Cu are relatively constant across the
film, and the ratios of Bi and Sr are consistent with the
stoichiometric ratios needed to form Bi2212.  For this deposition
the physical mask on Bi-Ca-2212 was changed to produce a constant
mass deposition across the film, while the Bi-Y-2212 target mask
produced a linear gradient across the radius of the doughnut
shaped sputtering track.  The result is an amorphous film with the
stoichiometric ratios needed to produce a set of
Bi$_2$Sr$_2$Ca$_{1-x}$Y$_x$Cu$_2$O$_y$ films with 0.1$<$x$<$0.4.
Post-annealing studies of the as-deposited conducted under a
variety of conditions will be reported in a subsequent paper.

\section{Conclusions}

Composition studies demonstrate the need for correct stoichiometry
across the whole film for the formation of the Bi2212 phase.
Through the development of a rotating carbon brush apparatus, a
-20 V DC bias can be applied to the sample substrate.  The
negative bias has been used successfully with an increased chamber
pressure to reduce the resputtering effect across the film.  Thus,
with these methods, one may produce the desired stoichiometry in
films prepared by the composition spread approach.

\section{Acknowledgements}
We would like to acknowledge valuable discussions with J. R. Dahn,
A. E. George and S. Trussler.  The financial support of the
Natural Sciences and Engineering Research Council of Canada is
gratefully acknowledged.


\begin{thebibliography}{00}

\bibitem{park95}
S. K. Park and J. H. Je, Physica C 245 (1995) 167.

\bibitem{hammond89}
R. H. Hammond and R. Borman, Physica C 162-164 (1989) 703.

\bibitem{westerheim91}
A. C. Westerheim, L. S. Yu-Jahes and A. C. Anderson, IEEE Trans.
Magn. 27 (1991) 101.

\bibitem{morris89}
D. E. Morris, C. T. Hultgren, A. M. Markelz, J. Y. T. Wei, N. G.
Asmar and J. H. Nickle, Phys. Rev. B 39 (1989) 6612.

\bibitem{grace92}
J.M. Grace, D.B. McDonald, M.T. Reiten, J. Olson, R.T. Kampwirth
and K.E. Gray, J. Vac. Sci. Technol. A 10 (1992) 1600.

\bibitem{grace91}
J.M. Grace, D.B. McDonald, M.T. Reiten, J. Olson, R.T. Kampwirth
and K.E. Gray, J. Appl. Phys. 70 (1991) 3867.

\bibitem{xu94}
J.-H. Xu, G.-G. Zheng, A.M. Grishin, B.M. Moon, K.V. Rao and J.
Moreland, Appl. Phys. Lett. 64 (1994) 1874.

\bibitem{nathan99}
S. S. Nathan, G. Mohan and S. Mohan, Thin Solid Films 347 (1999)
14.


\bibitem{blue91}
C. Blue and P. Boolchand, Appl. Phys. Lett. 58 (1991) 2036.

\bibitem{ballentine91}
P.H. Ballentine, J.P. Allen, A.M. Kadin and D.S. Mallory, J. Vac.
Sci. Technol. A 9 (1991) 1118.

\bibitem{klien91}
J.D. Klien and A. Yen, J. Vac. Sci. Technol. A 9 (1991) 1600.

\bibitem{cuomo78}
J.J Cuomo, R.J. Gambino, J.M.E. Harper, J.D. Kuptsis and J.C.
Webber, J. Vac. Sci. Technol. 15 (1978) 281.

\bibitem{kennedy65}
K. Kennedy, T. Stefansky, G. Davy, V. F. Zackay, E. R. Parker, J.
Appl. Phys. 36 (1965) 3808.

\bibitem{hanak70}
J. J. Hanak, J. Mater. Sci. 5 (1970) 964.

\bibitem{xiang95}
X. D. Xiang, X. Sun, G. Briceno, Y. Lou, K-A. Wang, H. Chang, W.
Wallace-Freedman, S-W. Chen and P. Schultz,  Science 268 (1995)
1738.

\bibitem{vanDover98}
R. B. van Dover, L. F. Schneemeyer, R. M. Fleming, Nature 162
(1998) 392.

\bibitem{chang99}
H. Chang, I. Takeuchi, X.-D. Xiang, Appl. Phys. Lett. 74 (1999)
1165.

\bibitem{xiang99}
X.-D. Xiang, Ann. Rev. Mater. Sci. 29 (1999) 149.

\bibitem{dahn02}
J.R. Dahn, S. Trussler, T.D. Hatchard, A. Bonakdarpour, J.N.
Meuller-Neuhaus, K.C. Hewitt and M. Fleischauer, Chem. Mater. 14
(2002) 3519.

\bibitem{hsu00}
I-J. Hsu, R.-S. Liu, J.-M. Chen, R.-G. Liu, L.-Y. Jang, J.-F. Lee
and K.D.M. Harris, Chem. Mater. 12 (2000) 1115.

\bibitem{rubin94}
L.M. Rubin, T.P. Orlando, and J.B. Vander Sande, Physica C 220
(1994) 284.

\bibitem{tsukamoto93}
K. Tsukamoto, H. Shimojima, M. Ishii, N. Enomoto and C. Yamagishi,
J. Am. Ceram. Soc. 76 (1993) 1031.


\end{thebibliography}
\end{document}